\documentclass[twocolumn,multicolumn,aps,prb,showpacs]{revtex4}
\usepackage{graphicx}
\usepackage{dcolumn}
\usepackage{bm}
\usepackage{color}
\usepackage{latexsym}
\usepackage[T1]{fontenc}
\usepackage{amsmath}

\input{epsf}

\begin{document}

\preprint{APS/123-QED}
\date{\today}

\title{Multifractal scaling of flux  penetration in the Iron-based Superconductor Ba(Fe$_{0.93}$Co$_{0.07}$)$_{2}$As$_2$}
\author{Mathieu Grisolia and Cornelis J. van der Beek}
\affiliation{Laboratoire des Solides Irradi\'{e}s, CNRS UMR 7642 \& CEA-DSM-IRAMIS, Ecole Polytechnique, F91128 Palaiseau cedex, France }
\author{Yanina Fasano }
\affiliation{Laboratorio de Bajas Temperaturas, Centro At\'{o}mico Bariloche \& Instituto Balseiro, Avenida Bustillo 9500, 8400 Bariloche, Argentina}
\author{Anne Forget,  Doroth\'{e}e Colson}
\affiliation{Service de Physique de l'Etat Condens\'{e}, Orme des Merisiers, CEA Saclay (CNRS URA 2464), 91191 Gif sur Yvette cedex, France}

\begin{abstract}
The penetration of magnetic flux fronts in the optimally-doped iron-based superconductor Ba(Fe$_{1-x}$Co$_{x}$)$_{2}$As$_2$ ($x = 0.07 \pm 0.005$) is studied by means of magneto-optical imaging and Bitter decoration. The higher-order analysis of roughening and growth of the magnetic flux front reveals anomalous scaling properties, indicative of non-Gaussian correlations of the disorder potential. While higher-order spatial correlation functions reveal multi-fractal behavior for the roughening,  the usual Kardar-Parisi-Zhang growth exponent is found. Both exponents are found to be independent of temperature. The scaling behavior is manifestly different from that found for other modes of flux penetration, such as that mediated by avalanches, suggesting that multi-scaling is a powerful tool for the characterization of roughened interfaces.
We propose a scenario for vortex penetration based on two-dimensional percolation and cluster aggregation for an inhomogeneously disordered superconductor.  
\end{abstract}

\maketitle

\section{Introduction}

Vortex line pinning and the ensuing irreversible magnetic properties of type-II superconductors have been studied for many years. These are usually described in terms of the Bean model and its generalizations.\cite{Bean62,Zeldov94,Brandt1996,Brandt98i,Brandt98ii,Mikitik2000,Mikitik2001}  While the Bean model accounts for global magnetic properties such as magnetization hysteresis loops and macroscopic flux distributions, it does not describe the local fluctuations of vortex densities $n_{v} = \langle{\mathbf B}\rangle_{\lambda_{L}} / \Phi_{0}$ in time and space,\cite{Surdeanu99,Surdeanu98} 
nor the roughness of the  magnetic flux penetration front. Here $\langle{\mathbf B}\rangle_{\lambda_{L}}$ is the coarse-grained flux density, averaged over a distance of the order of the penetration depth $\lambda_{L}$, and $\Phi_{0} = h/2e$ is the flux quantum. Due to the wide range of phenomena in which front growth and roughening occur, such as fluid flow in porous media,\cite{He92} propagation of the ignition front in burning paper,\cite{Surdeanu99,Maunuksela97} or the advancement of a rice pile,\cite{Pastor97,Aegerter2003,Aegerter2004,Lorincz2007,Denisov2012} and the many analogies between these different phenomena, the subject has raised a huge amount of  interest over the last 25 years. 

The analysis of  local variations $\delta h$ of the magnetic flux front height  $h(x)$ is commonly performed using the height-height correlation function\cite{Barabasi95}
\begin{equation} 
C^{2}\left( x,t\right) = \left\langle \left[ \delta h\left( x^{\prime },\tau \right) -\delta h\left( x^{\prime }+x,\tau +t\right) \right] ^{2}\right\rangle 
_{x^{\prime },\tau }
\label{Two point correlation function}
\end{equation}
where  $\delta h\left( x,t\right) = h\left( x,t\right) -\left\langle h\left( x,t\right) \right\rangle _{x}$ corresponds to the local deviation from the average front position, and $\left\langle ...\right \rangle _{x^{\prime },\tau }$ denotes averaging over the spatial coordinate as well as time. In our case, time steps  correspond to increments in the applied magnetic field $H_{a}$. 

The quantity (\ref{Two point correlation function}) enables one to simultaneously determine  both the roughness exponent $\alpha$ and the growth exponent $\beta$ by fitting the respective evolution
\begin{eqnarray}
C(x,0) \propto x^{\alpha} \hspace{2cm} (x \ll l_{sat}) \\
C(0,t) \propto t^{\beta} \hspace{2cm} (t \ll t_{sat}).
\end{eqnarray}
The saturation length $l_{sat}$ is the distance over which the front shape influences the height at a given site. The saturation time $t_{sat}$ is the time scale beyond which the influence of the previous evolution of the front is lost. 

Theoretical models predict  different scaling properties of a roughened interface (\em i.e. \rm different scaling exponents), depending on its (non)equilibrium state, the type of disorder to which it is subjected, and on its relaxational dynamics.\cite{Hagston99} Usually, one has a competition between two antagonistic mechanisms, such as the elasticity of the interface, which tends to smooth it, and its interaction with  a disorder potential, responsible for roughening.  A powerful theoretical approach to kinetic roughening is represented by the so-called Kardar-Parisi-Zhang (KPZ) model\cite{Kardar86} which describes the temporal evolution of the height variable $h(x,t)$, 
\begin{equation} 
\frac{\partial h}{\partial t} = \nu \nabla^2 h + \mu (\nabla h)^2 + \eta(x,h;t) + F.
\label{equation:KPZ}
\end{equation}
Here $\nu$ is an effective surface tension; $\mu$ quantifies the importance of lateral growth and vanishes for zero velocity, and $F$ is an external force. The disorder term $\eta(x,h;t)$ has a Gaussian distribution with zero mean, and $\langle \eta(x,t)\eta(x^{\prime},t^{\prime})\rangle = 2D\delta(x-x^{\prime})\delta(t-t^{\prime})$. Eq.~(\ref{equation:KPZ}) well describes diverse phenomena of kinetic roughening such as ballistic deposition or Eden models\cite{Kardar86} 
but also the advancement of flux fronts in type II superconductors.\cite{Surdeanu99} This is due to the growth of the front height being perpendicular to the front itself, since the Lorentz force on the flux vortices reads $F_{L}= \mathbf j \times \mathbf B$. The KPZ model predicts well-defined  roughness and growth exponents, $\alpha = \frac{1}{2}$ and $\beta = \frac{1}{3}$. In parallel, however, a wide variety of exponents 
has been reported in other systems, including wetting, imbibition, percolation, bacteria invasion, and so forth.\cite{Amaral94,Amaral95} Among different approaches to account for such phenomena, Barab\'{a}si {\em et al.}  proposed to apply the concept of multifractality to interface roughness.\cite{Barabasi91,Barabasi92,Santucci2007} Multifractality has been associated with a power-law distribution of the noise amplitude $\eta$, that accounts for rare events in the roughening or growing process. It is also relevant for the roughening of linear polymers on percolation clusters. \cite{Janssen2007,Blavatska2010,Janssen2012} Alternatively, a  multifractal formalism may be useful if different length scales compete in the growth process, or if depinning occurs at preferential sites.  In those cases, rare events such as avalanches may drive the kinetic properties.\cite{Leschhorn94,Paczuski96}

Regarding magnetic flux fronts in superconductors, Surdeanu {\em et al.}\cite{Surdeanu99} were the first to study different models to account for the roughening of the interface between the mixed state and Meissner phase  in thin--film cuprate superconductors. 
The authors\cite{Surdeanu99} distinguished two regimes in roughening and growth:  the short--range interaction regime, well-described by the Directed Percolation Depinning (DPD) model,\cite{Amaral95} 
and  the long--range regime, described by a usual 1+1-KPZ equation. By mapping the front progression through a quenched disorder potential on the problem of percolation, the DPD model predicts values $\alpha = 0.63$ and $\beta=0.65$. On the other hand,  Eq.~(\ref{equation:KPZ}) explicitly includes temporal fluctuations of the disorder. DPD-like behavior was also reported for flux penetration in Nb thin films on Si-substrates, over a narrow temperature window between smooth flux fronts at high $T$, and an avalanche-dominated regime at low $T$.\cite{Vlasko-Vlasov2004}  The authors\cite{Vlasko-Vlasov2004}  noted that the observed scaling might be due to KPZ behavior in the presence of spatially correlated (non-Gaussian) disorder. 

In this paper we characterize the penetrating flux front in single crystals of the optimally--doped  iron-based superconductor Ba(Fe$_{0.93}$Co$_{0.07}$)$_{2}$As$_2$ using the recently developed multi-scaling approach. The roughened front shows self-affine behavior, indicating non-Gaussian correlations of the disorder potential. Strikingly, the growth exponent is in agreement with the KPZ model. The multiscaling of the front roughness is clearly different from that produced by either avalanche dynamics in Nb thin films,\cite{Welling2004i,Welling2004ii,Aranson2005} or depinning  of a ferroelectric domain wall from a mixture of strong and weak pinning sites.\cite{Guyonnet2012} 

\begin{figure}[t]
\centering
\includegraphics[width=0.5\textwidth]{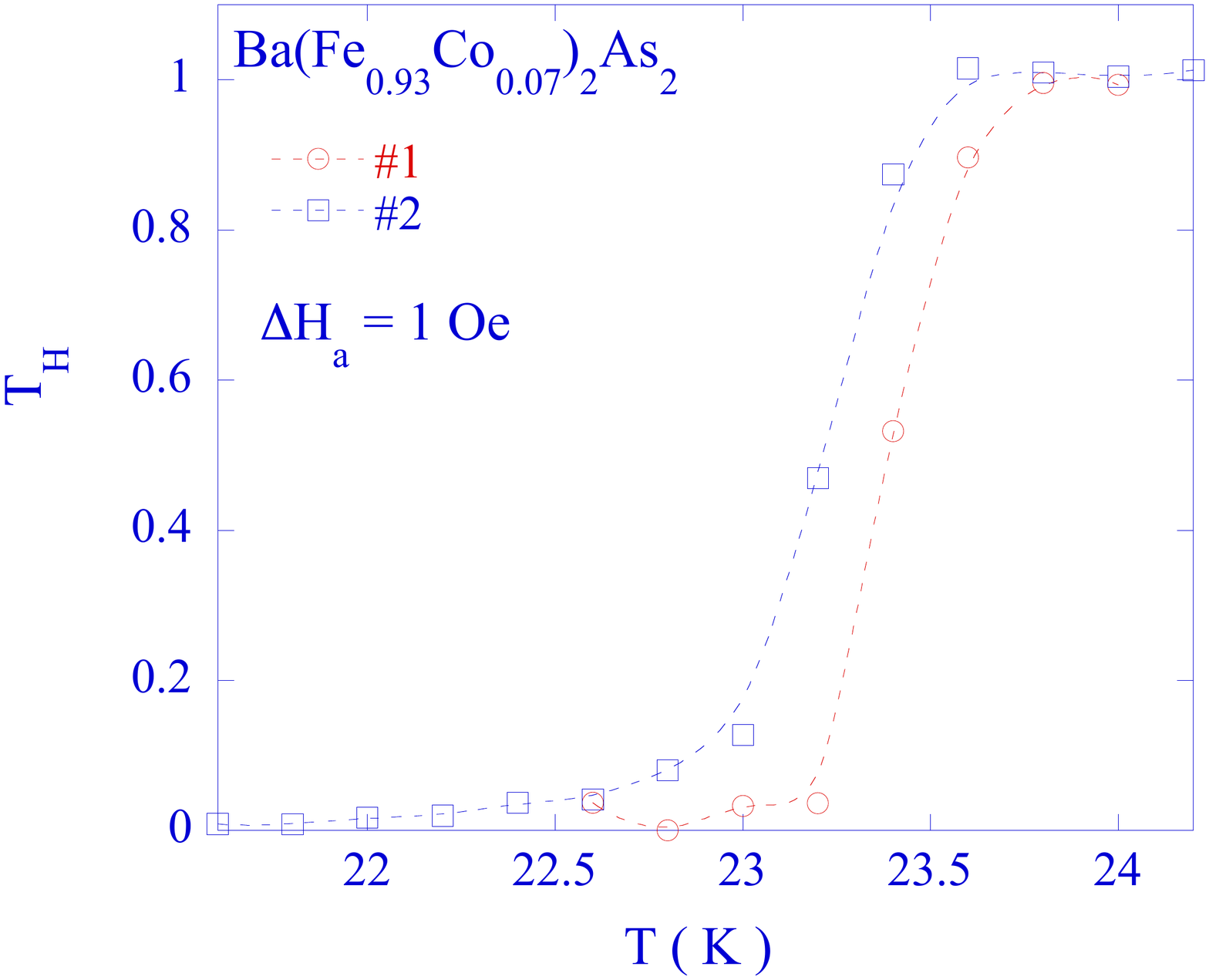}      
\caption{Superconducting transition of two crystals used in this study, as determined from differential magneto-optical measurements\cite{Demirdis2011} performed in $H=0$ with a field modulation of $\Delta H_{a} = 1$~Oe. Data are presented as the transmittivity $T_{H} \equiv [\overline{I}(T)-\overline{I}(T\ll T_{c})]/[\overline{I}(T>T_{c})-\overline{I}(T\ll T_{c})]$, where $\overline{I}$ is the averaged luminous intensity over the sample area. $T_{H} = 1$ and  $T_{H}=0$ correspond to the absence of screening and complete exclusion of magnetic flux respectively.}
\label{fig:TH}
\end{figure}   

\begin{figure}[t]
\centering
\includegraphics[width=0.35\textwidth]{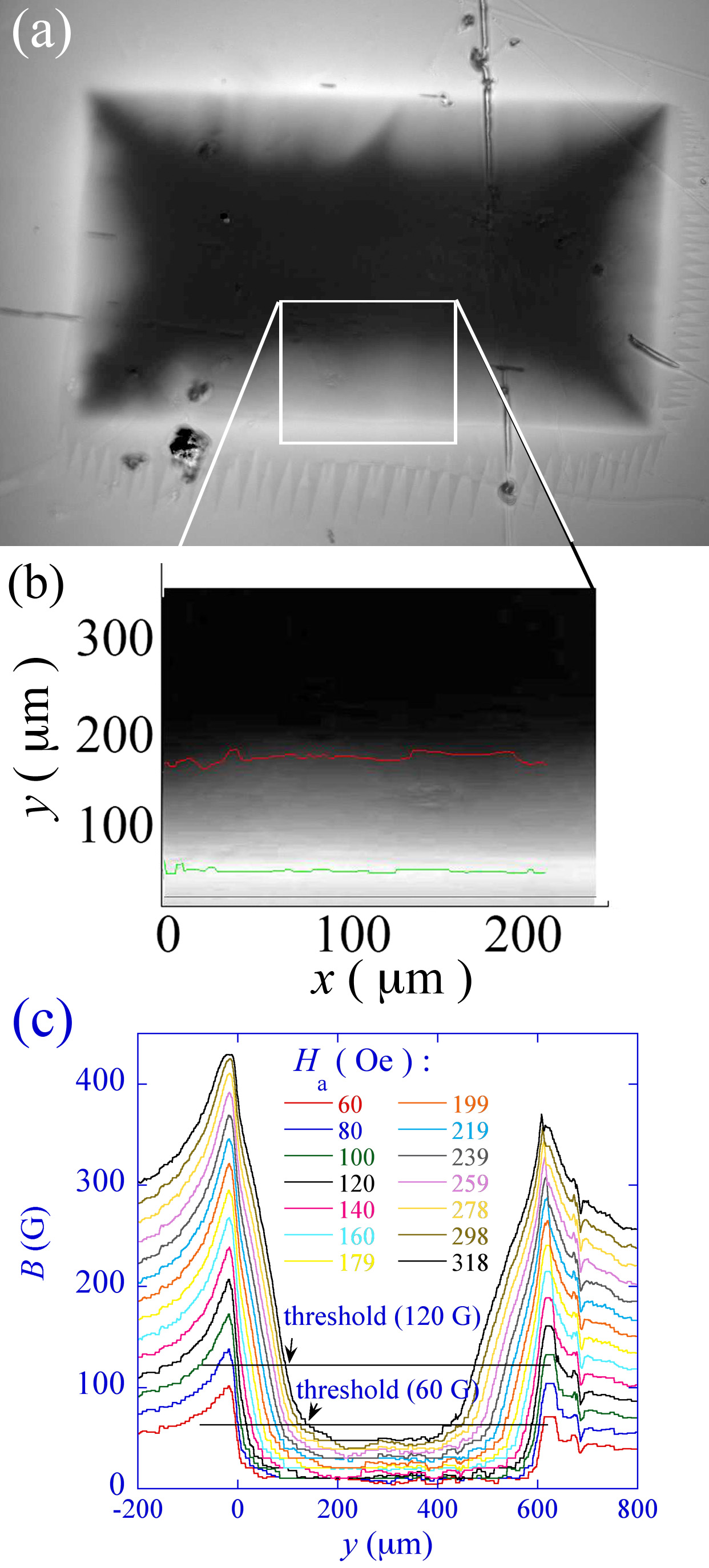}      
\caption{(a) Magneto-optical image of flux penetration into Ba(Fe$_{0.93}$Co$_{0.07}$)$_{2}$As$_{2}$ sample $\#$~1, taken at $H_{a} = 404$~Oe after zero-field cooling to 10~K.  (b) shows a zoom of the sample part studied in the analysis. The red line indicates the flux front,  determined as the induction threshold with $B = 120$~G. The green line shows the sample edge. (c)  Bean--like  profile of the average flux density, as a function of $y$, across the sample width. Flux density thresholds of 60 and 120 G are indicated.}
\label{fig:MOI}
\end{figure}   

We propose a tentative interpretation based on the findings of Ref.~\onlinecite{Demirdis2011} for Ba(Fe$_{1-x}$Co$_{x}$)$_{2}$As$_{2}$. That work reports that nanometer-scale inhomogeneity of the superconducting properties is at the origin of the lack of even intermediate--range vortex positional order, and of the significant vortex density fluctuations observed on field cooling.\cite{Demirdis2011} Spatial variations of superconducting parameters such as the critical temperature, $T_{c}$, or the superfluid density, $n_{s}$, would result in a random network of more-or-less favorable sites, suggesting the analogy with a percolation cluster. Independent evidence for such nanoscale disorder in Ba(Fe$_{1-x}$Co$_{x}$)$_{2}$As$_{2}$ was presented in Ref.~\onlinecite{Massee2009}.  The difference in multi-scaling of the roughness and growth exponents suggests the percolation and aggregation of different clusters (front sections) with different fractal dimension. 


\begin{figure}[t]
\centering
\includegraphics[width=0.6\textwidth]{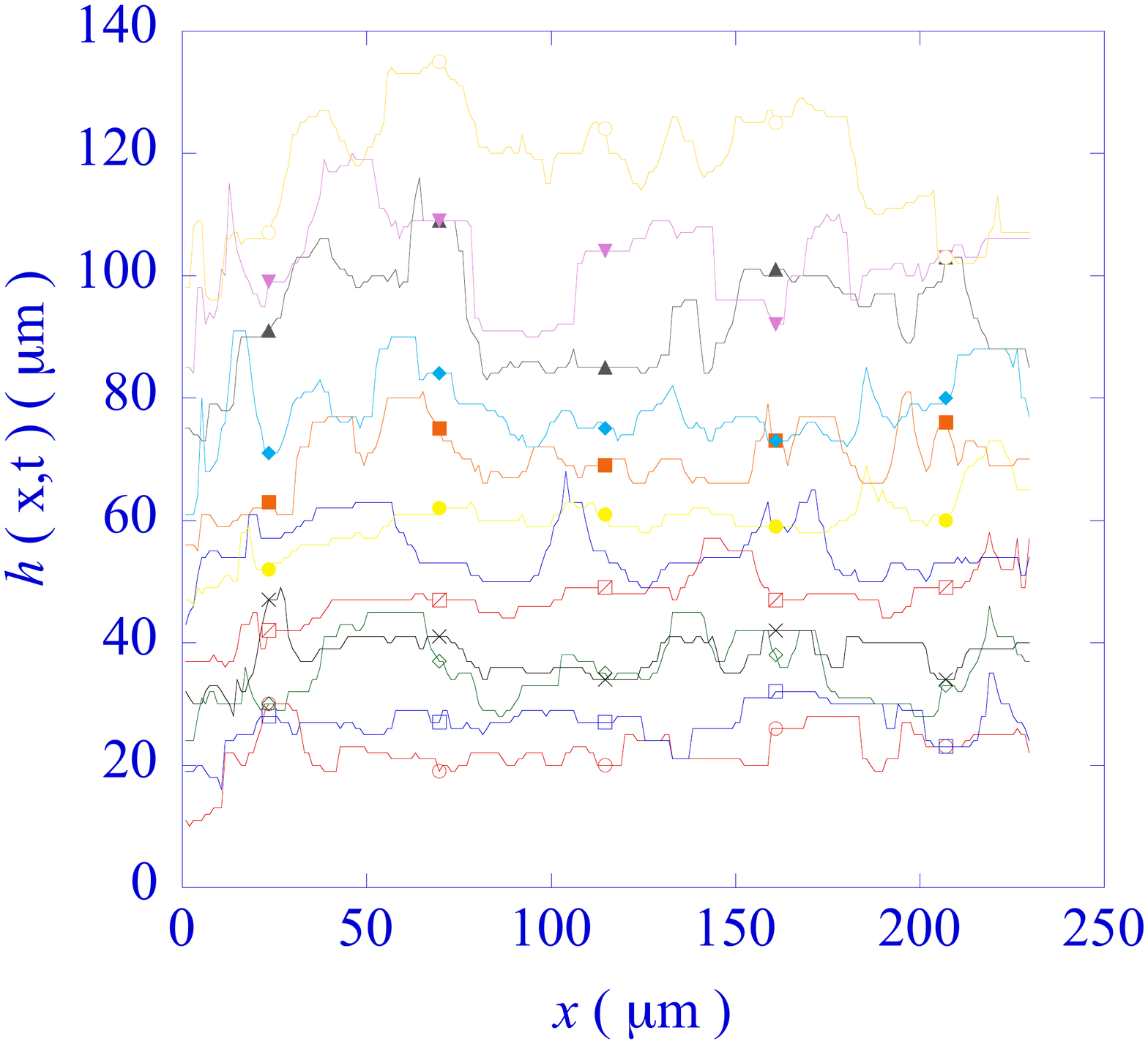}  
\caption{Flux fronts determined in sample \# 1 at 10 K, after zero field cooling, using an induction threshold of 120 G. The value of the external applied field ranges from 163 (bottom) to 383~Oe (top). }
\label{fig:front}
\end{figure} 

\section{Experimental Details}

Optimally--doped Ba(Fe$_{1-x}$Co$_{x}$)$_{2}$As$_2$ single crystals, with $x = 0.07 \pm 0.005$ and critical temperature $T_{c} = 23.8$~K, were grown using the self-flux method, as described in Ref.~\onlinecite{Rullier-Albenque2009}. The crystal composition was analyzed using a Camebax SX50 electron microprobe yielding the Co content within 0.5\% absolute accuracy. Rectangular samples  were cut from different crystals using a W wire saw (wire diameter 20~$\mu$m) and 1~$\mu$m SiC grit suspended in mineral oil. In particular, sample \# 1 has a length of 994~$\mu$m,  a width of  571~$\mu$m, and a thickness of 32~$\mu$m. Its critical current density (at $T = 10$~K)  is $j_{c}=2.4\times10^{8}\mathrm{ Am}^{-2}$. Sample \# 2  has a length of 835~$\mu$m, a  width of 733~$\mu$m, a thickness of 27~$\mu$m, and $j_{c}(10\,{\mathrm K}) = 3.1\times10^{8}\mathrm{ Am}^{-2}$. The penetration of magnetic flux into the selected samples was  visualized by the magneto-optical imaging (MOI) method.\cite{Dorosinskii92,Uehara2010} A ferrimagnetic garnet indicator film with in-plane anisotropy is placed on top of the sample. A non-zero perpendicular component of the magnetic induction $B_{}$ induces an out-of-plane rotation of the magnetization, and, thereby, a Faraday rotation of the polarization of light traversing the garnet.  An Al mirror evaporated on the hind side of the garnet reflects the impinging light, which is then observed using a polarized light microscope with nearly crossed polarizer and analyzer. Regions with nonzero $B_{}$ then show up as  bright when observed through the analyzer. The differential magneto-optical technique\cite{Demirdis2011} was used to characterize the superconducting transition. Fig.~\ref{fig:TH} shows that these are rather narrow for the material under study, of the order of 0.5~K.

Fig.~\ref{fig:MOI}(a) shows the penetration of magnetic flux into Ba(Fe$_{0.93}$Co$_{0.07}$)$_{2}$As$_{2}$ sample $\#$~1, the rectangular outline of which is clearly seen.  We shall be interested in the flux front within the sample, between peripheral bright regions of non-zero $B_{}$, and the dark central region of $B = 0$. This front corresponds to the interface between the mixed state,  in which the superconductor is penetrated by vortex lines, and the Meissner phase of excluded magnetic flux. In order to avoid bias and distortions induced by the effect of the sample corners on the shape of the flux front, we have studied the penetration near the center of the sample edges only.  At all measurement temperatures, the temperature stability was better than 10~mK; the external field was increased in 20~Oe steps in order to monitor the flux front progression. 
 
\begin{figure}[t]
\centering
\includegraphics[scale=0.14]{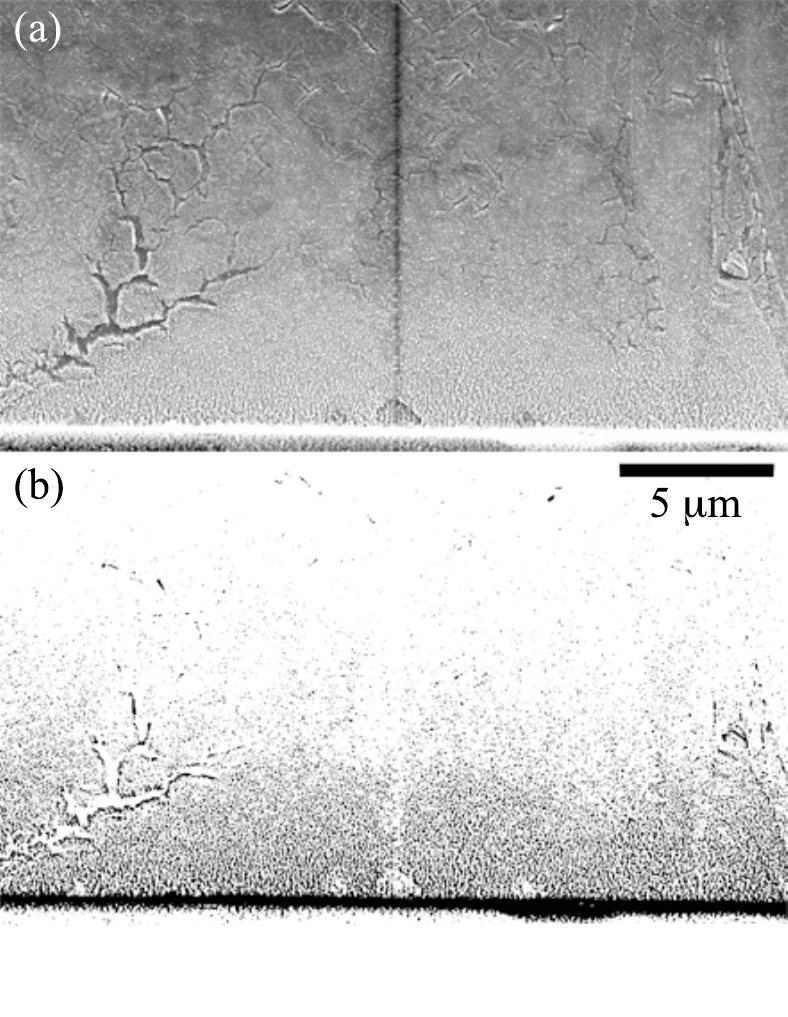}
\caption{(a) Bitter decoration image of flux penetration into sample \#~1, at an applied field $H_{a} = 50$ Oe (at which the mean distance between vortex lines $a_{0} = 0.63$~$\mu$m). The field is applied after zero-field cooling to 4.2~K. The image depicts the untreated Scanning Electron Micrograph of a 25 $\mu$m--long section of the region of the sample shown in Fig.~\protect\ref{fig:MOI}b. The sample edge corresponds to the bright line at the bottom. Individual vortices  appear as white dots. (b) \em ibid. \rm An intensity thresholding has been applied to the negative image of (a) in order to bring out the vortex positions as well as the flux front. The corrugation of the magnetic flux penetration front is clearly observed.}
\label{fig:Bitter}
\end{figure}

The MO images are converted to maps of the magnetic induction by calibrating the luminous intensity $I$ with respect to the applied field.\cite{Uehara2010} Next, the position of the flux front is determined from a given threshold level of the magnetic induction on the flux profiles. Using a non-zero threshold value avoids uncontrolled variations between experiments due to the specific  luminosity and polarization conditions under which images are acquired. In all, 17 different threshold values between 40 and 200 G were used.  The minimum threshold value that could be used for all applied field values was $B =120$~G. We define the height $h(x)$ of the flux front as the distance between the position at which the magnetic induction is maximum at the sample boundary ($y = 0$), and the position of the intensity threshold, along a 1 pixel-wide strip perpendicular to the sample boundary. Such a definition eliminates the effect of possible shifts in the luminous intensity due to over-exposure of certain strips in the camera. An example of the progression of the flux front  in sample \# 1 determined by this method is shown in Fig.~\ref{fig:front}. In order to evaluate the correlation function (\ref{Two point correlation function}), we subtract the mean position of the flux front $\langle h(t) \rangle_{x}$, averaged over the  250~$\mu$m--wide central section of the sample on which the analysis is performed.

In order to observe the flux front morphology on a finer scale, and, notably, to investigate the occurence of coalescing clusters, Bitter decoration\cite{fasano}  experiments were performed on the same samples after zero-field cooling to 4.2 K and the application of a field of 50 Oe.
Typical results (on sample \#1) are shown in Fig.~\ref{fig:Bitter}. The image shows individual vortex lines as these enter the crystal from the lower edge. Note that the  intervortex distance near the edge $a_{0} \equiv (\Phi_{0}/\left\langle B \right\rangle_{\lambda_{L}})^{1/2} = 0.63$~$\mu$m. Similar images were obtained along all four sample edges.


\begin{figure}[t]
\centering
\includegraphics[width=0.61\textwidth]{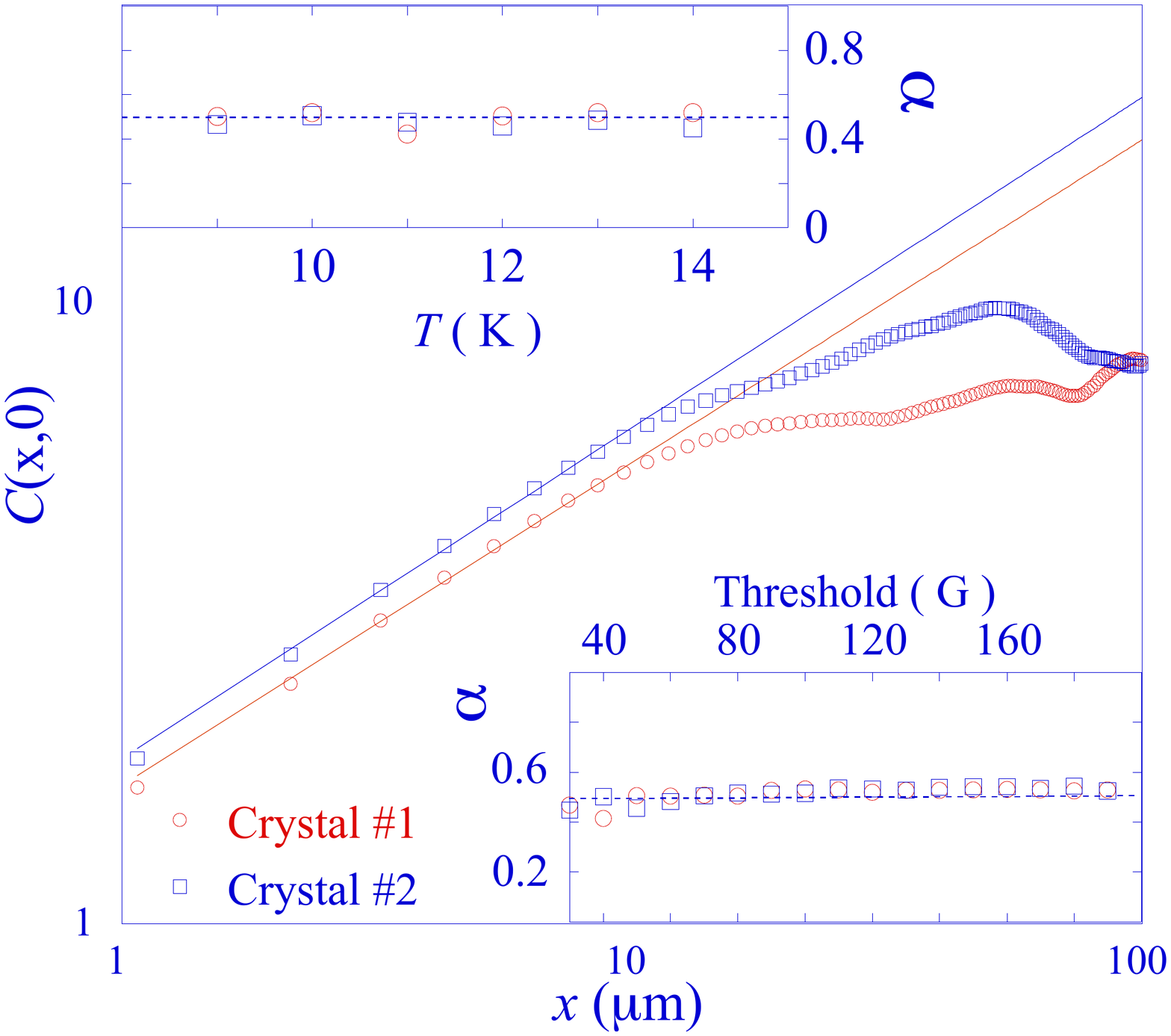}      
\caption{Main panel: Two-point correlation function of the penetrating flux front, {\em i.e.} the square-root of the quantity of Eq.~(\protect\ref{Two point correlation function}), calculated for the flux front at 10~K in samples \#~1(red circles 
) and \# 2 (blue squares 
). The roughness exponent is, by definition, the slope of the linear (solid lines) regime of $C(x,0)$ in a log-log plot. Its value is in very nice agreement with the KPZ model, {\em i.e.} $\alpha =  0.50$. Upper inset: Temperature independence of the roughness exponent. Lower inset:  Independence of $\alpha$ on the value of the induction threshold, as this ranges from 40 to 200 G. In both insets,  the dashed lines indicate $\alpha \approx  0.5$.}
\label{fig:c(x,0)}
\end{figure} 
%

\begin{figure}[t]
\centering
\includegraphics[width=0.55\textwidth]{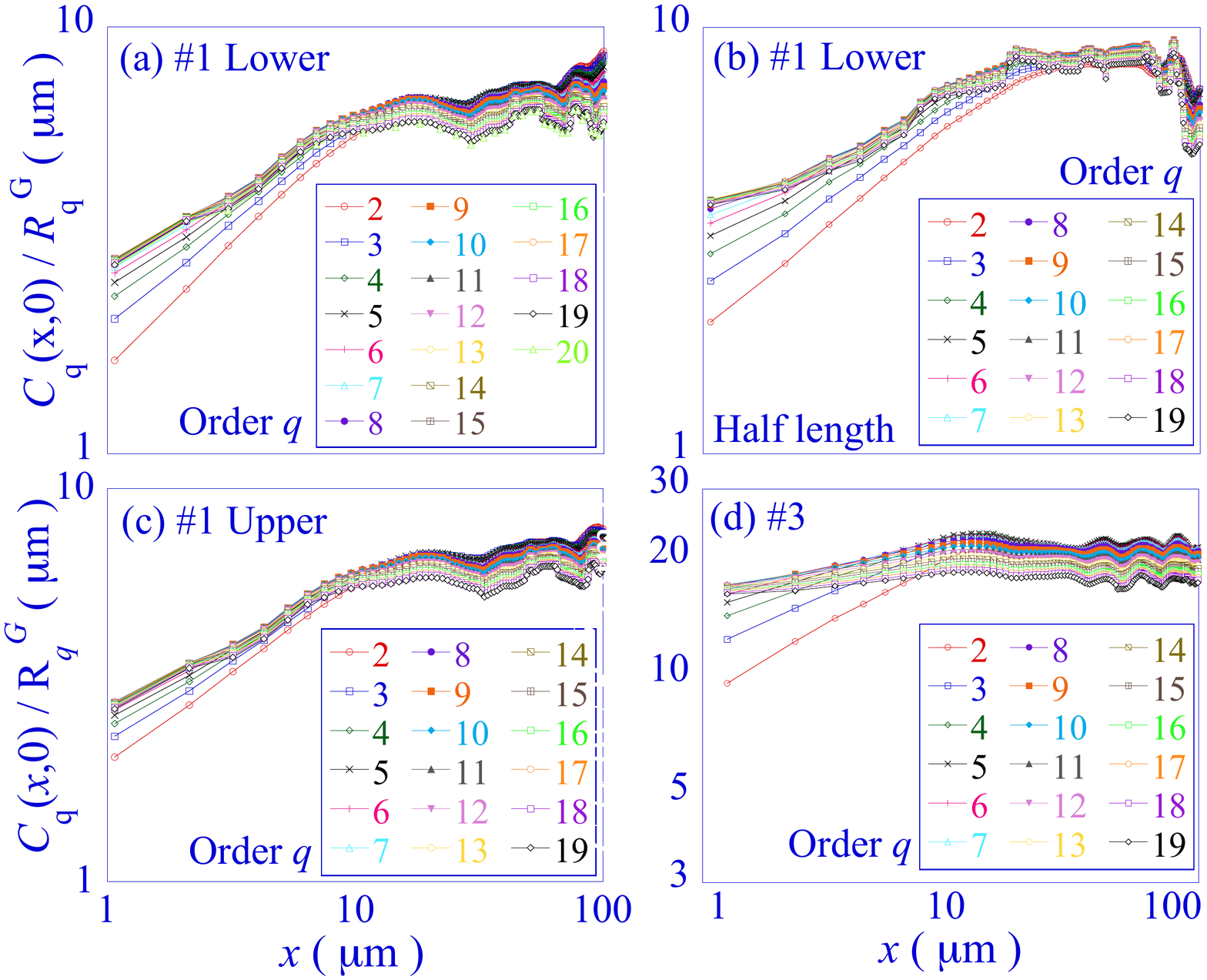}      
\caption{Higher order spatial correlation function $C_q(x,t)$, normalized by the Gaussian ratios $R_{q}^{G}$,\cite{Santucci2007,Guyonnet2012} with $q$ varying from 2 to 19, as measured on the penetrating flux front in Ba(Fe$_{0.93}$Co$_{0.07}$)$_{2}$As$_2$ sample \# 1 and  \#~3  at $T = 10$~K. (a) Sample \#~1, front at lower edge, such as imaged in Fig.~\protect\ref{fig:MOI}; (b) {\em ibid.}, but only half the front length is taken into account; (c) sample \#1, front at upper edge; (d) sample \#~3. In all investigated cases, the slope depends continuously on $q$ and presents nontrivial scaling. The behavior is therefore multifractal. }
\label{fig:manyorder}
\end{figure} 

\section{Analysis of the Flux Front}

The main panel of Fig.~\ref{fig:c(x,0)} shows the spatial correlation function $C(x,0)$, describing the roughness of the flux front penetrating from the lower edge of sample \#~1 at $T = 10$~K.  The value of $\alpha$ determined from the logarithmic slope of $C(x,0)$ versus $x$ is very close to the value derived in the KPZ model, $\alpha = \frac{1}{2}$.\cite{Barabasi95} In this respect, our data are similar to those of Ref.~\onlinecite{Surdeanu99}. The constant value of $C(x,0)$ for $x_{sat} \gtrsim 10$~$\mu$m indicates that deformations of sections of the flux front separated by more than 10~$\mu$m are independent. The value of $x_{sat}$  is possibly related to the (similar) dimensions of independently penetrating vortex clusters, to be discussed below. 

In order to ascertain the robustness of these results, we have checked for temperature and vortex density dependence. 
At sufficiently low flux densities the distance $a_{0}$ between vortex lines  exceeds $\lambda_{L} (T)$.  Since $\lambda_{L}$ increases with $T$, the interaction between vortices also increases with $T$ within this $B$-range (for a fixed vortex density). On the contrary, at higher 
vortex densities $a_{0} < \lambda_{L}(T)$, and the repulsive interaction decreases with $T$. Hence, vortex lines are not  sensitive to the same length scales of disorder for different $T$ and $B$, so that, for non-trivial correlations of the disorder, a continuous evolution of $\alpha(T,B)$ might be expected.  However, the upper inset to Fig.~\ref{fig:c(x,0)} clearly shows that the roughness exponent does not depend on temperature.  

Investigating the dependence of the roughness exponent on vortex density corresponds to probing its dependence on the threshold value.  The constant $\alpha$ as function of  threshold, displayed in the lower inset to Fig.~\ref{fig:c(x,0)},  is consistent with the lack of dependence on temperature, since both imply that the front roughness is insensitive to the strength of the vortex interaction. In other words, the scaling properties of the flux front will depend only on the kind of disorder and on the manner in which the front deformations are relaxed.

 
\begin{figure}[t]
\includegraphics[width=0.65\textwidth]{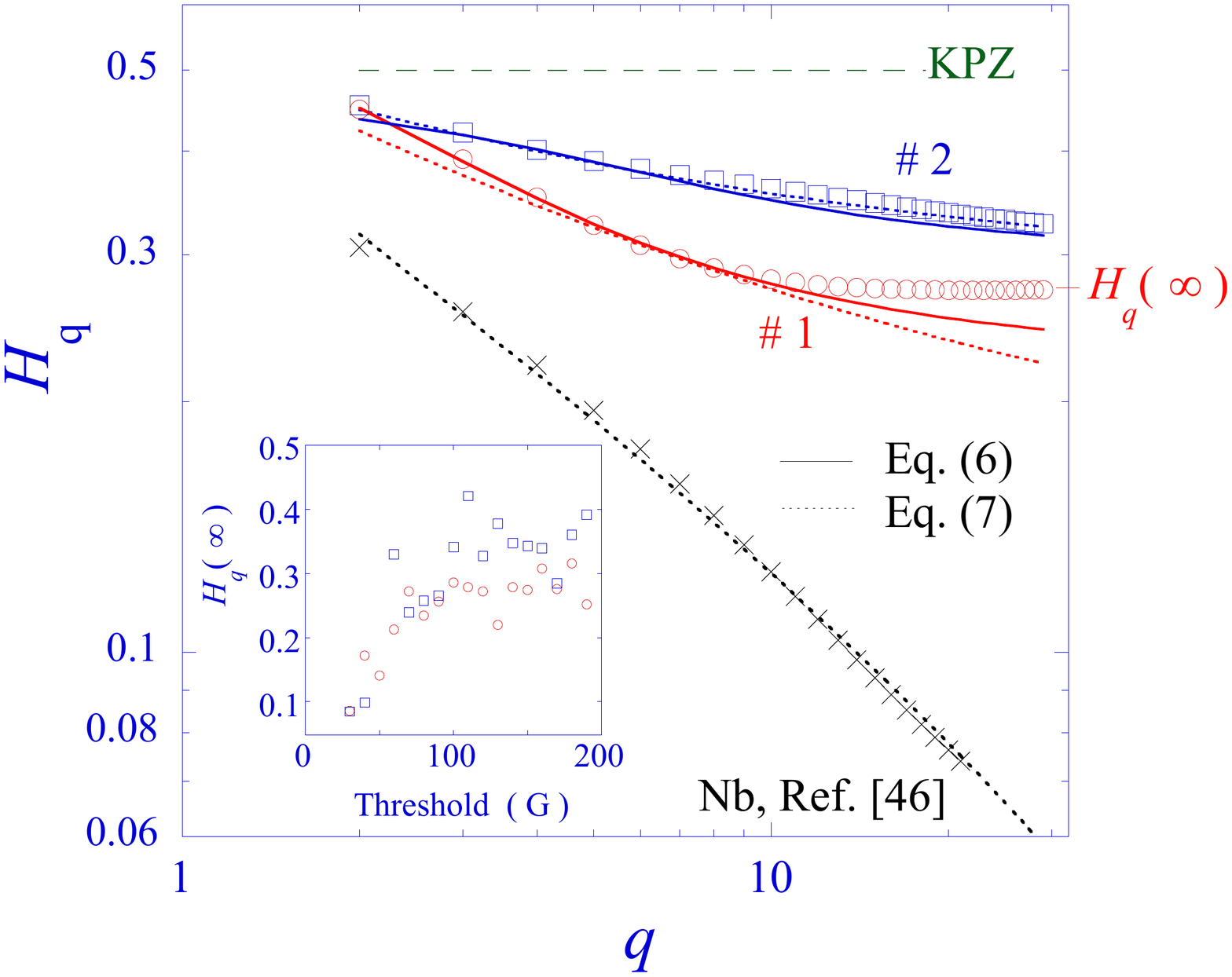}          
\caption{The exponent $H_{q}$ as a function of the order $q$ for the magnetic flux front in Ba(Fe$_{0.93}$Co$_{0.07}$)$_{2}$As$_2$ samples \# 1 (red circles) and \# 2  (blue squares), evaluated for a threshold of $B = 120$~ G. For comparison, $H_{q} = \frac{1}{2}$ for the KPZ model is also shown (green diamonds), as is the result of the analysis of the flux penetration observed at 6~K in the superconducting Nb films of Refs.~\protect\onlinecite{Welling2004ii} and \protect\onlinecite{Aranson2005}  ($\times$). The decay of $H_{q}$ measured on the superconducting samples is monotonous, without any inflection. For the Ba(Fe$_{0.93}$Co$_{0.07}$)$_{2}$As$_2$ samples, $H_{q}$ saturates at a lower, resolution--limited bound $H_{q}(\infty)$. The evolution of $H_{q}(\infty)$ as function of the flux-density threshold is shown in the Inset. $H_{q}(q)$ compares well with Eq.~(\protect\ref{eq:toy-model}), with ($b_{1},b_{2}) = (0.2, 0.73)$ and  $(0.4,0.66)$ for samples 1 and 2, respectively. A comparison with Eq.~(\protect\ref{eq:Duplantier}) (dotted lines) yields parameters sets $(a,b,c) = (2,26.3,0.15)$, $(1.2,24,0.28)$, and $(2.1,17.6,-0.03)$ for samples 1,  2, and the Nb--film of Refs.~\protect\onlinecite{Welling2004ii} and \protect\onlinecite{Aranson2005}, respectively. }
\label{fig:Hq}
\end{figure} 

We now turn to the multiscaling analysis of the data. The method\cite{Santucci2007,Blavatska2010,Guyonnet2012} relies on the computation of the higher-order two--point correlation function
\begin{equation}
C_q(x,t)={\langle \left[\delta h(x^{\prime},\tau)-\delta h\left(x^{\prime}+x,\tau + t\right)\right]^q\rangle_{x',\tau}}^{1/q}. 
\label{Many order two-point correlation function} 
\end{equation}
Following previous work,\cite{Barabasi91} $C_q$ should scale as  $C_q(x,0) \propto x^{H_q}$, with $H_{q}$ the generalized Hurst exponent. A non-trivial $q$-dependent scaling is the hallmark of a non-Gaussian probability density function (PDF) of the disorder, leading to a self-affine (but not self-similar) interface.\cite{Barabasi91} More generally, it may indicate multifractal behavior, {\em i.e} the geometry of the interface is not simply fractal but  implies many different geometries of many different fractal dimensions.\cite{Duplantier99}  On the contrary, a constant $H_{q} = \alpha = 0.5$ is representative of KPZ behavior and of an interface with a Gaussian PDF (see Ref.~\onlinecite{Barabasi92} and Fig.~\ref{fig:Hq}). Figure~\ref{fig:manyorder} shows the higher order two-point correlation functions evaluated on the flux fronts in sample \# 1 and \#~3,  normalized by the factors $R_{q}^{G} \equiv C^{G}_{q}(x,0)/C^{G}_{2}(x,0)$,\cite{Santucci2007,Guyonnet2012}  with $q$ ranging from 2 to 20. Here, the correlation functions $C_{q}^{G}(x,t)$ are those that would be obtained for an interface with a Gaussian PDF of the local displacements.\cite{Santucci2007,Guyonnet2012}  We note that the same multiscaling is observed even when a subsection of the investigated front [Fig.~\ref{fig:manyorder}(a) and (b)] or  a different edge of the sample is considered, see Fig.~\ref{fig:manyorder}(c). Figure~\ref{fig:Hq} shows the behavior of the exponent $H_{q}$ as a function of $q$ for the magnetic flux fronts in our Ba(Fe$_{0.93}$Co$_{0.07}$)$_{2}$As$_2$ crystals.  $H_{q}$ is observed to saturate to a sample--dependent value $H_{q}(\infty)$ at large $q$. From Figs.~\ref{fig:manyorder} and the exponent $H_{q}$ rendered in \ref{fig:Hq}, it is clear that, for small $x$, the scaling with length depends on the order $q$ in a non-trivial manner. The analysis of different sections of the same flux fronts yields an identical $H_{q}(q)$--dependence. The $H_{q}(q)$-evolution on the opposite side of the same sample, and for flux fronts in the other investigated samples is similar in character, but not identical. 

The $q$-dependence of the Hurst exponent describing the flux fronts in Ba(Fe$_{0.93}$Co$_{0.07}$)$_{2}$As$_2$ is  remarkably well-rendered by the simple toy model used in Ref.~\onlinecite{Barabasi91} to prove the relevance of multi-scaling for the description of self-affine fractal interfaces,\cite{Barabasi91}
\begin{equation}
H_{q} = \frac{\ln \left[ \left( b_{1}^{q} + b_{2}^{q} \right) / 2 \right]}{q \ln \frac{1}{4}}.
\label{eq:toy-model}
\end{equation}
Here $b_{1}$ and $b_{2}$ are the scaling parameters characterizing the self-affine interface structure parallel and perpendicular to the growth direction, such as these are used in Ref.~\onlinecite{Barabasi91} for the construction of a model interface. The experimental values, ($b_{1},b_{2}) = (0.2, 0.73)$ and  $(0.4,0.66)$ for samples \#~1 and \#~2, respectively, clearly demonstrate the irrelevance of a Gaussian PDF for the transverse excursions of the flux front in Ba(Fe$_{0.93}$Co$_{0.07}$)$_{2}$As$_2$, which would yield a constant roughness exponent $\frac{1}{2}$. A self-similar interface with Gaussian PDF can be generated in the toy model\cite{Barabasi91} by using  $b_{1} = b_{2} = 0.5$.

\begin{figure}[t]
\centering
\includegraphics[width=0.6\textwidth]{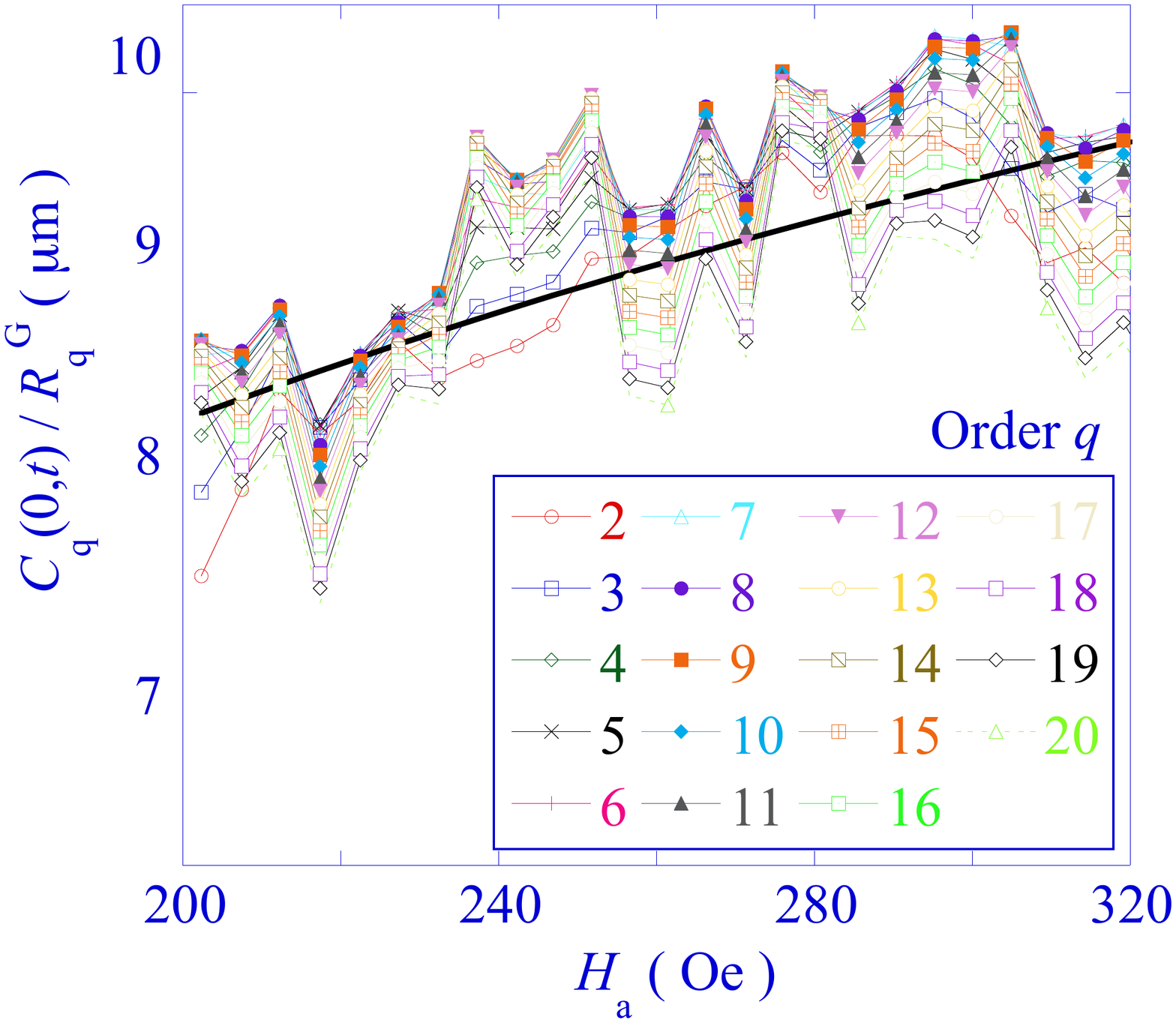}      
\caption{Higher-order two-point correlation function as a function of the temporal evolution of the magnetic flux front in Ba(Fe$_{0.93}$Co$_{0.07}$)$_{2}$As$_2$ sample \# 1, at $T = 11$~K. Data, normalized using the Gaussian factors $R_{q}^{G}$,\cite{Santucci2007,Guyonnet2012} are plotted as a function of the externally applied magnetic field (as the temporal variable). For the calculation of these curves, only field values between 197 and 393~G were taken into account -- for lower fields, flux does not penetrate, for higher fields, complete penetration is achieved. The $C_{q}(0,t)$ show the same behavior for all $q$, ranging from $q = 2$  to 20, and scale with the $R_{q}^{G}$--values. The data compare well with the KPZ prediction, $C_{q}(0,t) \sim t^{1/3}$(black line).  }
\label{fig:higherordergrowthq}
\end{figure}

We have also evaluated multiscaling of growth of the flux front height. Figure~\ref{fig:higherordergrowthq} shows that the temporal correlation functions $C_{q}(0,t) = C_{q}(0,H_{a}/\dot{H}_{a})$ are parallel for all $q$, and scale with the Gaussian factors $R_{q}^{G}$ ($\dot{H}_{a}$ is the rate at which $H_{a}$ is increased). The temporal correlation functions obey the same power-law behavior,  $C_{q}(0,t) \sim  t^{0.33}$. We consider this value to be in agreement with $\beta = \frac{1}{3}$ from KPZ theory. We thus find, surprisingly, that the evolution of the flux fronts in the Ba(Fe$_{0.93}$Co$_{0.07}$)$_{2}$As$_2$ samples corresponds to the diffusive growth of a multi-fractal structure.

\begin{figure}[b]
\centering
  \includegraphics[width=0.5\textwidth]{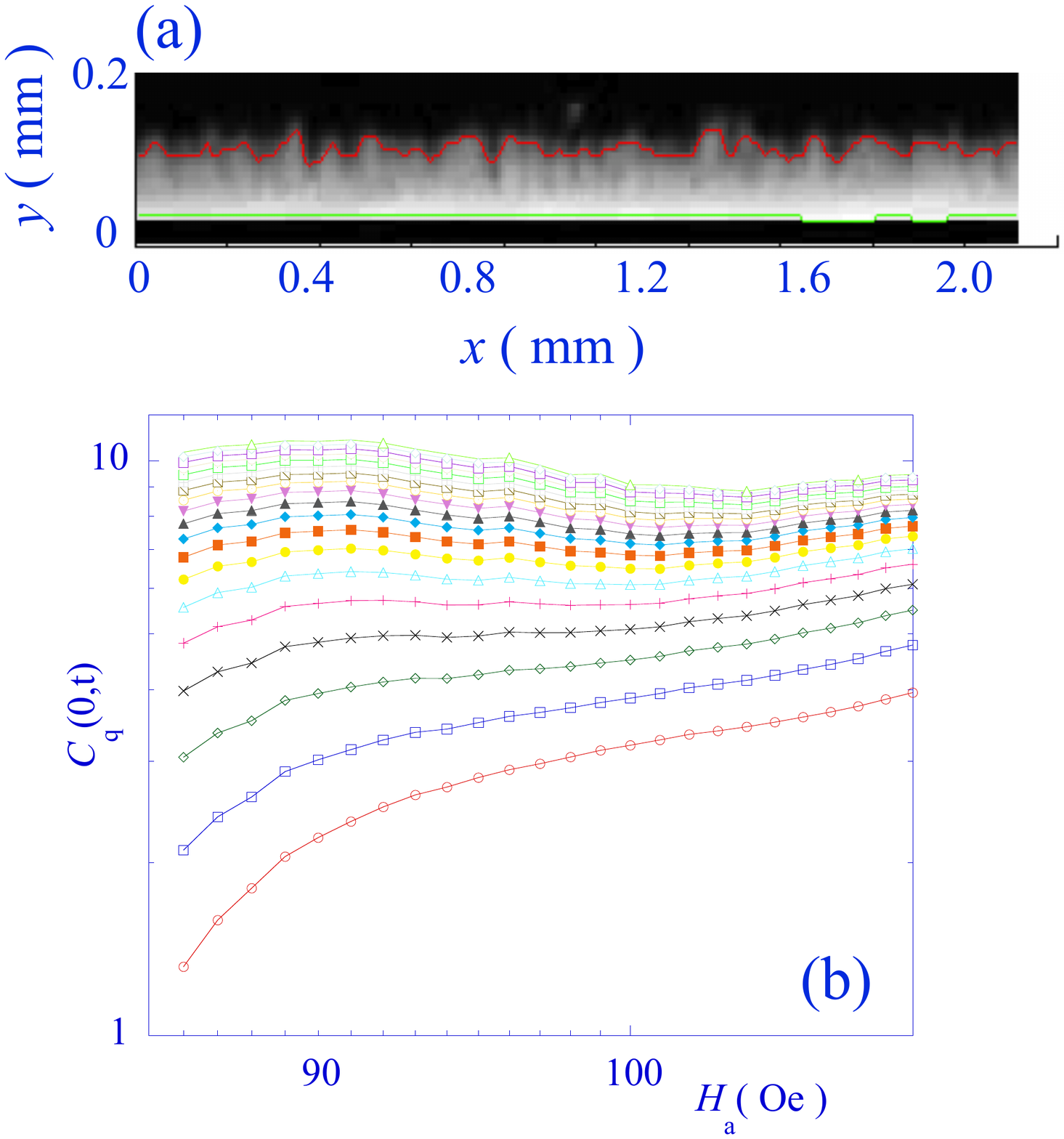}         
 \vspace{-38mm}
\caption{(a) Flux front in 500 nm-thick Nb thin film, deposited on A-plane sapphire, at $T = 6$~K and $H_{a} = 1.5$Oe (data taken from Ref.~\protect\onlinecite{Supplemental}
). (b)  Multi-scaling analysis of the growth behavior of such a flux-front. Contrary to the case of Ba(Fe$_{0.93}$Co$_{0.07}$)$_{2}$As$_{2}$,  $C_{q}(0,t)$ also has a $q$-dependent temporal behavior. The $q$--dependent correlation function fans out at small times, {\em i.e.} small applied fields, but shows $q$-independent behavior at large times.}
\label{fig:Aransongrowth}
\end{figure}

\section{Discussion}

A number of propositions have been made to explain the origin of multifractality and multiscaling. A non-Gaussian, {\em e.g.} power--law, PDF underlying the disorder term $\eta$ was introduced for the description of percolative imbibition of paper by a fluid.\cite{Buldyrev92} The link between penetrating flux fronts and the porous medium equation for the imbibition by fluids was previously pointed out in Refs.~\onlinecite{Surdeanu99} and \onlinecite{Gilchrist94}. However, the porous medium equation as such, as analyzed in Ref.~\onlinecite{Gilchrist94}, does not lead to (multi--) fractal behavior. 

A second  explanation for anomalous scaling in roughening processes is the occurrence of avalanches.\cite{Paczuski96,Bassler98,Bassler99,Aegerter2003,Aegerter2004} The numerical work of Bassler {\em et al.}  showed the possible multi-fractal character of braided vortex trajectories resulting from avalanches in a strongly pinning superconductor.\cite{Bassler99} The effect of local temperature variations on the nucleation and propagation of dendritic vortex avalanches was investigated experimentally by Welling {\em et al.},\cite{Welling2004i,Welling2004ii} who studied thin superconducting Nb films deposited on sapphire substrates of various orientations. The authors\cite{Welling2004i,Welling2004ii} found that, in 500 nm-thick Nb deposited on A--plane sapphire, flux penetration at low temperature ($T < 6$~K) takes place via huge compact avalanches, while at $T > 6$~K it is more regular, yielding continuous flux fronts. The latter bear a certain resemblance to those observed in the present work. Numerical work by Aranson {\em et al.}\cite{Aranson2005} suggests that avalanches in defect-free films occur at periodic locations, while avalanches in films with edge defects are initiated at these imperfections. With this in mind,  it was suggested\cite{Aranson2005} that both avalanche--like and more regular flux penetration in Nb/A--plane sapphire is initiated by edge defects. In order to ascertain whether the roughened flux fronts in Ba(Fe$_{0.93}$Co$_{0.07}$)$_{2}$ have a similar origin, we have analyzed the structure of the continuous flux fronts  ($T > 6$~K) reported in 
movie 8 of Ref.~\onlinecite{Supplemental}, which shows flux penetration such as that reproduced in Fig.~\ref{fig:Aransongrowth}(a). To perform this analysis, we considered a threshold of 10~\% of the applied field.  
The analysis of the correlation function (\ref{Many order two-point correlation function}) shows that the flux fronts in this Nb thin film show multi-scaling behavior (Nb data in Fig.~\ref{fig:Hq}). The Hurst exponent starts off with a lower value $H_{2} =  0.3$, and decreases as a near power--law to zero as $q$ increases.  Contrary to our results on Ba(Fe$_{0.93}$Co$_{0.07}$)$_{2}$As$_2$, this curve is not satisfactorily described by Eq.~(\ref{eq:toy-model}). Moreover, the high-temperature flux penetration as shown in Ref.~\onlinecite{Welling2004ii},  \onlinecite{Aranson2005} , and \onlinecite{Supplemental} \em  does \rm obey multi-scaling during growth, see Fig.~\ref{fig:Aransongrowth}(b). An explanation for this is that even at temperatures higher than 6~K, the ``regular'' flux penetration in Nb thin films would be the result of the superposition of many avalanches. The observed $q$-dependence would then be induced by the avalanche size distribution, which follows a power--law distribution in space and time. 

Clearly, the roughening and growth of the flux front observed in Ba(Fe$_{0.93}$Co$_{0.07}$)$_{2}$As$_2$ crystals is dissimilar from the superposition of thermo-magnetic avalanche-like instabilities as observed in the case of Nb thin films. A more pertinent analogy may be that suggested by the analysis of roughened ferroelectric domain walls.\cite{Guyonnet2012} Ref.~ \onlinecite{Guyonnet2012} reports on ``mono-affine'' scaling of the domain-walls, with a Gaussian PDF of its local transverse displacements, on small length scales, both in numerical simulations as in actual experiments on Pb(Zr$_{0.2}$Ti$_{0.8}$)O$_{3}$. Such a mono-affine behavior corresponds to so-called ``weak-collective pinning'', in which the domain walls are only pinned by the fluctuations in the pin density.  Analysis on larger length scales showed ``multi-affine scaling'' described by a non-trivial hierarchy of higher-order scaling exponents. The crossover length scale in the study\cite{Guyonnet2012} was given by that on which rare events in the form of strongly pinning defects, such as dislocations, occur. Translated to our experiments, this would correspond to the mono-affine (KPZ-like) growth of individual front sections emanating from specific sections of the boundary of the superconducting crystal, separated by sections from which flux penetration is less likely. Such edge inhomogeneity might then be interpreted in terms of the local variation of superconducting properties\cite{Demirdis2011} and / or geometrical irregularities at the sample edges. However, in our analysis multiscaling is found to be the same irrespective of considering the entire length of the flux front (see Fig.~\ref{fig:MOI}(b)) or a subsection thereof (see Fig.~\ref{fig:manyorder}).  

A third possibility is the growth and coalescence of front sections with different fractal geometry, arising from different disorder realizations in distinct parts of the Ba(Fe$_{0.93}$Co$_{0.07}$)$_{2}$As$_2$ crystals. Such heterogeneity may be expected, {\em e.g.}, from the analysis of the highly disordered vortex ensembles observed using Bitter decoration.\cite{Demirdis2011} The interaction--energy distribution of these ensembles could be understood only if one admits the presence of substantial spatial heterogeneity of the superconducting parameters, on the scale of 10\,--\,100 nm. Front sections with different roughness characteristics would then show diffusive growth, according to the KPZ description, but form an anomalously roughened front upon coalescence. Coalescence was considered in the numerical study of linear polymers on percolation clusters by Blavatskaa and Janke.\cite{Blavatska2010} They proposed that  multiscaling could arise from the merger of two fractal structures of different dimensions,  keeping their underlying geometry in the process.\cite{Blavatska2010} Once a cluster is formed, its temporal roughening is characterized by a certain growth exponent; the cluster retains the same geometry and dimension during growth. This behavior could account for both the observed temporal monoscaling growth behavior and the multifractal roughening upon the collision of two roughened structures. Indeed, Fig.~\ref{fig:Bitter} shows evidence for percolative penetration of vortices at the very interface of the mixed state and the Meissner state in Ba(Fe$_{0.93}$Co$_{0.07}$)$_{2}$As$_2$. Vortex lines are seen to penetrate the sample in an irregular fashion, the flux front on the $\mu$m--scale featuring peninsula--like protrusions separated by vortex--free areas. These features then coalesce to generate a continuous front. 

A  description of both the coarse-graining of the flux front as one goes from individual vortex-resolution (Fig.~\ref{fig:Bitter}) to the magneto-optical images of Fig.~\ref{fig:MOI}, and the non-trivial $H_{q}\left( q \right)$-dependence of the coarse-grained flux--front is suggested by the results of Ref.\onlinecite{Duplantier99}. The author\cite{Duplantier99} considered percolative growth in two dimensions, both for random and self-avoiding walks. The $q$--dependence of the exponent $\alpha(q)$ describing the metric of the harmonic measure of a two--dimensional near-critical percolation cluster is given by
\begin{equation}
\alpha(q) =   c  + \frac{a}{\sqrt{b q+1}},
\label{eq:Duplantier}
\end{equation}
with $a = \frac{5}{2}$, $b = 24$, and $c = \frac{1}{2}$ [Eq.~(5) of Ref.~\onlinecite{Duplantier99}]. The harmonic measure of a percolation cluster  is characterized by a lower dimension than the cluster itself, due to the inaccessibility of sites situated on deep ``fjords'' on the latter. Aggregation at certain sites on the percolation cluster would correspond to the growth of the flux front at specific locations due to the arrival of a vortex line. Eq.~\ref{eq:Duplantier}  very well describes the experimental $H_{q}(q)$--dependence of Fig.~\ref{fig:Hq}, including that for the flux front in Nb thin films, for very similar parameter values. This would imply that the flux-front in heterogeneous type-II superconductors can be described as the  hull of a near-critical percolation network.

\section{Summary and Conclusions}

We have measured and analyzed magnetic flux-penetration fronts in single crystals of the optimally--doped iron based superconductor Ba(Fe$_{0.93}$Co$_{0.07}$)$_{2}$As$_{2}$, over a wide range of temperatures. Analysis reveals multi-scaling of the  higher--order two-point spatial correlation functions of the roughened flux front. This implies that the roughness of the  front cannot be described by simple diffusive behavior, that is, by disorder with a Gaussian probability density function. By implication, the multi-scaling approach is a powerful tool to distinguish 
between the different properties at the origin of interface roughening.  Scaling of the flux-front roughness does not depend on temperature or the induction threshold used to define the front position, nor on macroscopic defects that may exist in particular samples. In contrast, a regular KPZ-like growth of the flux-front is found, excluding avalanche--like behavior as being at the origin of the front roughening.  We propose an interpretation of our results in terms of multi--fractal roughening due to the aggregation of vortex clusters with various fractal dimensions. Such clusters could in fact be identified using Bitter decoration, which reveals the structure of the flux front on the scale of individual vortex lines.  Once the macro cluster is formed, the front develops in time ({\em i.e.} with increasing magnetic  field) with a classical KPZ exponent.
We tentatively decribe this unusual roughening  by a theory for the harmonic measure of a two-dimensional percolation hull.\cite{Duplantier99}  Our results underscore the analogy between percolation in porous media and vortex penetration in inhomogeneous superconductors.

\section*{Acknowledgements} This work was partially funded by the grant  ``MagCorPnic'' of the R\'{e}seau Th\'{e}matique de Recherche Avanc\'{e}e   ``Triangle de la Physique'' du Plateau de Saclay,  by the Agence Nationale de la Recherche grant "PNICTIDES", and by the support of the ECOS-Sud-MINCyT France-Argentina bilateral program, Grant No. A09E03. Work done in Bariloche was partially funded by PICT 2007-00890 and PICT 2008-294.


\end{document}